# Anisotropic Surface Plasmon Polariton Generation Using Bimodal V‑Antenna Based Metastructures

Daniel Wintz,[†] Antonio Ambrosio,[†,‡] Alexander Y. Zhu,[†] Patrice Genevet,[†,§] and Federico Capasso[*,†]

[†]Harvard John A. Paulson School of Engineering and Applied Sciences, Harvard University, 9 Oxford Street, McKay 125, Cambridge,
Massachusetts 02138, United States
[‡]CNR-SPIN U.O.S. Napoli, Dipartimento di Fisica, Università di Napoli Federico II, Complesso Universitario di Monte Sant'Angelo,
Via Cintia, 80126 - Napoli, Italy
[§]Centre de Recherche sur l'Hétéro-Epitaxie et ses Application, CNRS, Rue Bernard Gregory, Sophia-Antipolis, 06560 Valbonne,
France

ABSTRACT: V-shaped nanoantennas are among the popular choices for the unit element of a metasurface, a nanostructured surface used for its ability to mold and control the wavefront of light. In general, the motivation for choosing the V-antenna as the unit element comes from its bimodal nature, where the introduction of the second mode offers extra control over the scattered wavefronts. Here, through near-field scanning optical microscopy, we study a 1D metastructure comprised of V-antennas in the context of generating asymmetric surface plasmon polariton (SPP) wavefronts. The key point is that the use of the V-antenna allows for the creation of a two-dimensional phase gradient with a single line of antennas, where the extra phase dimension offers additional control and allows for asymmetric features. Two different asymmetries are created: (1) SPP wavefronts that have different propagation directions on either side of the metastructure, and (2) SPP wavefront asymmetry through focusing: one side of the metastructure focuses SPP wavefronts, while the other side has diverging SPP wavefronts.

KEYWORDS: surface plasmon polaritons, metamaterials, nanoantennas, metasurfaces, near-field scanning optical microscopy.

Surface plasmon polaritons (SPPs) are electromagnetic modes that exist as propagating surface waves bound at the interface of a metal and dielectric. SPP modes offer an alternative to free space propagation, subwavelength energy confinement, and exhibit novel optical phenomena.[1] As SPPs are "dark modes" confined to a surface, directly imaging SPPs is difficult and requires probing the near-field component of the electromagnetic field, scattering the energy into a far-field propagating mode. This is most commonly achieved through near-field scanning optical microscopy (NSOM),[2,3] where a scanning probe tip is brought into the near-field and either scatters the SPPs into free space modes (apertureless, scattering-type NSOM) or into a propagating mode in an optical fiber that is then sent to a detector (aperture, collection mode NSOM). Metasurfaces,[4–6] with their ability to tailor the phase of light on a subwavelength scale, are particularly suited for engineering wavefronts, resulting in optical components such as lenses,[7,8] holograms,[9–12] and other exotic optical elements.[13–15] SPP devices based on the phased-array methodology of metasurfaces have also resulted in the creation of SPP lenses,[16–19] unidirectional couplers,[20–22] and holography based devices.[23–25] The design of any optical metastructure is based on two things: the unit element that imparts a phase shift to the incident light, and the phase profile of the desired optical component. In this work, we distinguish two separate cases of asymmetric scattered wavefronts: (1) wavefronts with different propagation directions on either side of the metastructure, and (2) wavefronts that focus or defocus on either side of the metastructure. To achieve such control, we utilize the Babinet inverse of a V-antenna, a nanoaperture in the shape of a V as our unit element. V-antennas have received a lot of interest in the field of metasurfaces, including many early works, and have been characterized both in the far-field[4,5,26] and near-field.[27] A single V-antenna supports two orthogonal electric dipole modes excited by orthogonal polarizations of light, a symmetric mode and an antisymmetric mode, where the symmetric or antisymmetric refers to the charge oscillation about the axis of the V. It is the existence of the two modes that offers an extra degree of control of the wavefront scattered from the Vantenna relative to a single mode antenna. As such, the bimodal or dual resonance methodology has become commonplace in the field of metasurfaces.[28,29] As shown in ref [26], both the resonant frequency and the scattering cross section of each mode can be independently controlled by tuning the length of the antenna arms and the angle of the V-shaped antennas. In our case, the extra dipolar mode of the V-antenna is used to introduce a two-dimensional phase gradient (in both x and y) with a single line of phased antennas (1D metasurface), enabling us to achieve asymmetric SPP wavefronts. In order to understand the role of the V-antenna in creating asymmetric wavefronts, it is instructive to describe how a twodimensional phase gradient can induce asymmetry in a wavefront. As a simplest case and proof of concept, consider a line of hypothetical point sources of SPPs, displaced from each other by an amount $\Delta x$ in the x direction and $\Delta y$ in the y direction. Determining the resulting wavefronts from the line of SPP sources is a matter of calculating a phase matching condition. Figure 1 shows the geometry of the SPP point sources with relevant parameters highlighted. There are two parts that contribute to the phase of the SPPs: (1) the propagation phase, equal to $k_{SPP}\Delta L$, where $k_{SPP}$ is the SPP wavevector, and $\Delta L$ is the path length difference between two sources, and (2) a phase difference inherent to the sources themselves (i.e., different initial phases), which we denote as $\Delta\phi$ (the origin of this term is a consequence of the metastructure and is known as the Pancharatnam-Berry phase, which will be discussed later). From a geometric analysis of Figure 1 and some algebra (see Supporting Information), the resultant angles of the wavefronts can be derived:

$$\sin \gamma_1 = \frac{1}{k_{SPP}}\frac{\Delta\phi}{\Delta x} + \frac{\Delta y}{\Delta x}\cos \gamma_1$$

$$\sin \gamma_2 = \frac{1}{k_{SPP}}\frac{\Delta\phi}{\Delta x} - \frac{\Delta y}{\Delta x}\cos \gamma_2$$

where $\gamma_1$ and $\gamma_2$ are the wavefront angles, $k_{SPP}$ is the SPP wavevector, $\Delta\phi/\Delta x$ is the Berry phase induced phase gradient in the x direction, and $\Delta y/\Delta x$ is the slope of the line that connects the elements, which would be zero for a one dimensional phase gradient. We note that the wavefront angles appear on both sides of the equation, requiring numerical analysis to solve. Setting $\Delta y/\Delta x$ to zero recovers the symmetric wavefront angle equation from ref [30]. The asymmetry in the wavefront angles is a consequence of the sign the $\Delta y/\Delta x$ term. In ref [30], a 1D metastructure consisting of sequentially rotated line aperture antennas was used to create symmetric SPP wavefronts, where the wavefront angles were studied and analyzed via NSOM. An analogy was made to "wakes", a wave phenomenon that occurs when a disturbance travels through a medium at a velocity faster than the phase velocity of the waves it creates, causing a characteristic buildup that explains the physics behind wakes from boats, Cherenkov radiation, and sonic booms. By choosing a suitable 1D metastructure, the authors were able to create a running wave of polarization along the metastructure with a phase velocity that exceeded the phase velocity of the SPPs, thereby creating SPP "wakes". Instead of using a 1D metastructure consisting of linear antennas, V-antennas allow us to design a two-dimensional phase gradient using a one-dimensional array of rotated antennas. Figure 2 illustrates the two modes of the bimodal V-antenna. The symmetric mode (Figure 2a) is symmetric about the axis of the V-antenna and can be approximated by a point dipole situated at the center of the V-antenna. The antisymmetric dipole mode (Figure 2b) is asymmetric about the axis of the V-antenna and can be approximated by a point dipole at the midpoint of the line connecting the two ends of the arms of the V. Given that the effective length of the antenna that supports charge oscillation is different for the two modes (2L for the antisymmetric mode and L for the symmetric mode), the modes have different resonance wavelengths. We use finite difference time domain (FDTD) simulations (see Supporting Information) to obtain the scattering amplitude (resonance curve) for a single antenna as a function of wavelength for a given angle $\Delta$ and arm length L. We choose the parameters (angle $\Delta$ and arm length L) such that both modes have equal scattering amplitudes at a free space operating wavelength of $\lambda_0 = 631$ nm. Figure 2c,d highlights two different arrangements of sequentially rotated V-antennas. By design, the symmetric mode center always lies on the x axis, whereas the antisymmetric mode, displaced from the symmetric mode, is not constrained to lie on the x axis. This is a key point: the antisymmetric mode traces out a two-dimensional path and is responsible for the two-dimensional phase gradient. Figure 2c shows a distribution of antisymmetric mode centers that lie only below the x axis, while Figure 2d shows an even distribution of the antisymmetric mode about the x axis. We note that according to FDTD, there is a small scattering amplitude due to the quadrupole mode associated with the asymmetric charge oscillation (as in Figure 2b). We incorporate this into the antisymmetric mode as an effective dephasing of the mode, relative to the symmetric mode. The symmetric mode and antisymmetric mode will be displaced spatially by an amount L/2 cos(Δ/2). To account for the dephasing due to the quadrupole mode, we simply let the geometric separation between the modes vary as a fitting parameter, to be deduced from the experimental data. Thus, the dephasing due to the quadrupole mode is incorporated as a propagation phase, similar to ref 22. This methodology will also effectively incorporate dephasing due to fabrication imperfections as well. The origin of the phase gradient is known as the Pancharatnam-Berry phase.[31] It is a geometric phase accumulated by illuminating sequentially rotated V-antenna apertures with circularly or elliptically polarized light. The x and y components of circularly polarized light are dephased from each other, and this dephasing causes a dephasing in the excitation of the symmetric and antisymmetric modes: this explains why the modes are excited $\pm \pi/2$ out of phase, depending on the handedness of the polarization. Furthermore, by sequentially rotating the V-antennas, each V-antenna can be dephased from the previous antenna by an amount equal to the rotation.

Consider only the antisymmetric mode centers in Figure 2d. If the period of the elements is much bigger than the displacement of the antisymmetric mode from the x axis, the sinusoidal distribution can be well approximated by a line. Hence, eqs 1 and 2 will accurately describe the wavefront angles. However, the arrangement shown in Figure 2c cannot be approximated by a single line, but by two lines of equal and opposite slope (forming a triangle with the x axis). The y direction phase gradient induced by these two lines will cancel out, leaving only the phase gradient in x to describe the wavefront angles, effectively recovering the one-dimensional case. In order to understand the behavior of these bimodal antennas, we analytically simulated SPP electric field distributions of such arrangements of bimodal structures. We approximate each mode of each antenna as a point dipole, with a phase dependent on: the mode (symmetric or antisymmetric, excited $\pi/2$ out of phase), and the rotation of the antenna (Berry phase term). Then, the fields of all the simulated dipoles are added together, employing the superposition principle to get the resultant total field. Furthermore, these fields are also added to electric field of the incident Gaussian beam transmitting through the gold, resulting in interference and allowing us to view the interference fringes of the SPPs and the incident light. This is similar to the actual experimental near-field imaging conditions. The simulation results for different cases are presented in Figure 3. Figure 3a shows the SPP electric field distribution for only the symmetric mode, which lies solely on the x axis. This configuration only has a one-dimensional phase gradient, and thus has symmetric wavefronts with angles given by eq 1 or 2 with Δy set to 0. Figure 3c shows only the antisymmetric mode (to isolate the effect of the 2D phase gradient), with a rotation arrangement shown by Figure 2d. As can be seen, this arrangement introduces asymmetry into the wavefront angles given by eq 1 and 2. However, if the one-sided arrangement of Figure 2c is used for the antisymmetric mode, as can be seen in Figure 3b, little to no asymmetry in the wavefront angles is recovered because the phase gradient in the y direction cancels out. In this case, a focusing effect is seen in the wavefronts below the structure, which is due to the dipoles tracing out a curve. Flipping the handedness of the incident light polarization does not change the side of focusing because this feature only results from the symmetric and antisymmetric dipole positions.

To perform experiments, we first e-beam evaporated 50 $\mu$m of gold onto a polished silicon substrate, 1 $\mu$m thick. We then template strip the gold from the silicon wafer,[32] using UV curable epoxy (Norland Optics 61) and a glass slide. This is done to reduce the surface roughness of the gold and increase the SPP propagation lengths. Next, we take the gold on glass and use focused ion beam (FIB, Zeiss NVision) to mill the Vantenna apertures into the gold film, with a beam current of 10 pA, with the resulting metastructure shown in the scanning electron micrograph of Figure 4a. Next, we mount our sample into a near-field scanning optical microscopy setup. Laser light at 632 nm is incident on the sample from below (transmitting through the glass and gold), and an NSOM tip scans the top surface containing the metastructure. The NSOM tip is comprised of an atomic force microscopy tuning fork operating in tapping mode with a tapered optical fiber with a diameter of ∼150 nm attached. In principle, collecting signal from the nearfield allows for subdiffraction limit imaging, but this is not necessary for our experiment. While scanning the surface in tapping mode, the NSOM tip scatters the SPPs and transmitted laser light into a propagating mode in an optical fiber, which is sent to a single photon avalanche photodiode to record the signal. Figures 4 and 5 present near-field scanning optical microscopy data for the focusing 2D gradient (Figure 2c) and the asymmetric wavefronts 2D gradient (Figure 2d), respectively. Figure 4b−d shows the behavior of the structure under different polarizations of the illuminating light. In Figure 4c, the 2D focusing effect is most noticeable because there is no phase gradient accumulated along the structure under linearly polarized light, isolating the geometrical effect of the dipole displacements. We find a focal distance from the data: f = 2.5 ±0.6 $\mu$m, in reasonable agreement with the theoretical value of 2.0 $\mu$m (see Supporting Information). The large uncertainty is due to interference・experimental fringes are spaced by $\lambda_{SPP}$; the focal distance cannot be known more accurately than $\lambda_{SPP}$. Figure 5 shows NSOM data from an array of V antennas with a higher scattering cross section for the antisymmetric mode, allowing us to highlight the asymmetric wavefront angles effect. As mentioned earlier, it is beyond the scope of this paper to know the exact effect of fabrication imperfections on the individual V-antenna size (i.e., the ion beam mills wider arms, etc.) and the effect of the antisymmetric quadrupolar mode on the dephasing between the modes. Thus, we estimate from our data that the dephasing between the modes translates to an effective modal separation of $\lambda_{SPP}/6.8$, that is, the positions of the blue and red dots in Figure 2a,b are effectively separated by $\lambda_{SPP}/6.8$ due to dephasing effects. In this paper, we have studied and shown the effects of using a bimodal unit element in a metastructure, which can be used to induce various anisotropies, namely, focusing or anisotropic effective indices for isotropic media. We use near-field scanning optical microscopy to experimentally measure and verify the induced anisotropies. We demonstrate that this extra mode and 2D phase gradient can be harnessed to create and control asymmetric surface plasmon wavefronts. As another example of the possible anisotropies, we demonstrate asymmetric focusing/defocusing on either side of such a metastructure. We anticipate this work being used to design more complex and innovative metasurfaces, where the extra control of manipulating a second mode in the unit element can be used for more exotic wavefront control.

## AUTHOR INFORMATION

Corresponding Author
*E-mail: dwintz@seas.harvard.edu.
ORCID
Daniel Wintz: 0000-0001-9948-1948

Funding
The authors acknowledge support from the Air Force Office of Scientific Research under Grant FA9550-14-0389 (MURI). P.G. is supported by the European Research Council (ERC) under the European Union's Horizon 2020 research and innovation program (Grant Agreement FLATLIGHT No.639109).
Notes
The authors declare no competing financial interest.

## ACKNOWLEDGMENTS

The authors acknowledge the Harvard Center for Nanoscale Systems (CNS), which is a member of the National Nanotechnology Infrastructure Network (NNIN). The authors acknowledge Nanonics Ltd. and Horcas et al.[33] for support in the near-field setup and WSxM software, respectively.

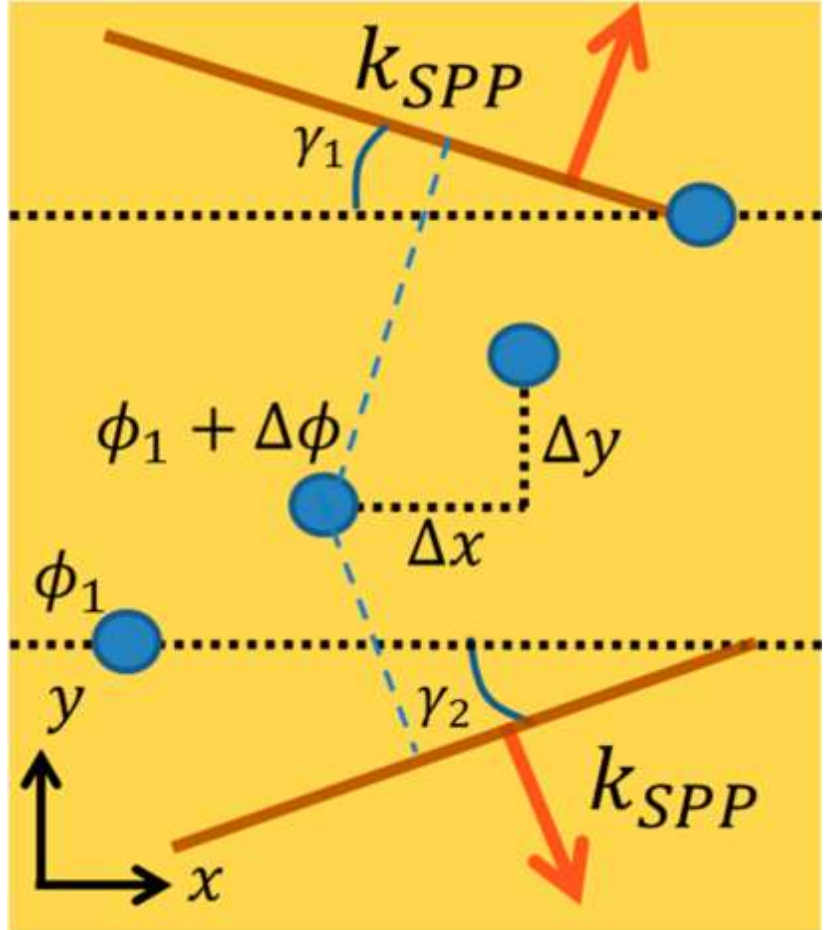

**Figure 1.** 2D phase gradient geometry. Schematic illustrating how a phase gradient in two dimensions results in the formation of wavefronts (brown lines) with different angles, $\gamma_1$ and $\gamma_2$. Surface plasmon polariton (SPP) sources (represented by blue circles), are offset in two dimensions by amounts $\Delta x$ and $\Delta y$. The sources have phases $\phi_1$ and $\phi_1 + \Delta\phi$, respectively. Calculating the wavefront angles amounts by setting the phase accumulation for the two sources to be equal: $k_{SPP}(L_1 + L_2) = \Delta\phi$ for $\gamma_1$ and $k_{SPP}(L_3) = \Delta\phi$ for $\gamma_2$, with results shown in Equations 1 and 2.

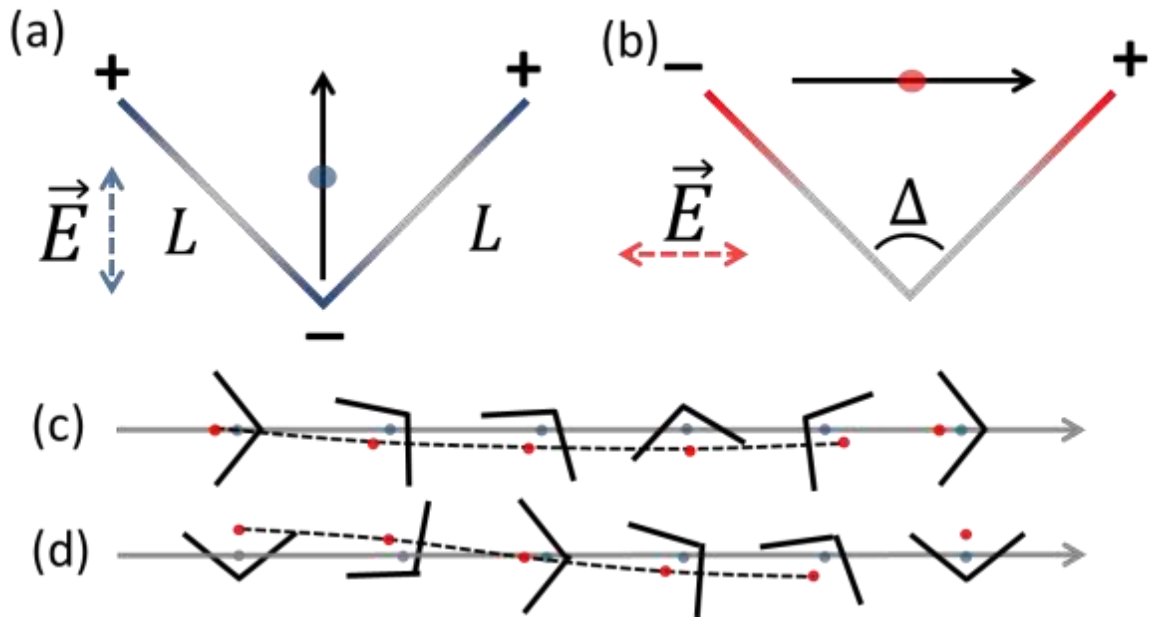

**Figure 2.** Bimodal V-antennas. (a) Symmetric mode of the V-antenna of arm length *L*, excited by light polarized in the direction given by the blue dashed arrow. The mode can be approximated by a single dipole, centered at the blue dot, with dipole moment given by the black arrow. (b) Same as (a), but for the antisymmetric mode, excited by light polarized in the direction of the red dashed arrow. This mode can be approximated by a single point dipole centered at the red dot, with dipole moment given by the black arrow. Note that the dipole centers of (a) and (b) are displaced from each other in space. (c) and (d) depict two different arrangements of rotated V-antennas, with the red and blue dots related to the positions of the antisymmetric and symmetric modes, respectively. The black dashed lines highlight the spatial movement of the antisymmetric mode throughout the period, denoting two different types of two-dimensional phase gradients that can be imposed, where (c) highlights an uneven distribution of the antisymmetric mode centers (with respect to the midline, denoted by the grey arrow) and (d) highlights an even distribution. The center of the symmetric mode of each V-antenna always lies on the midline.

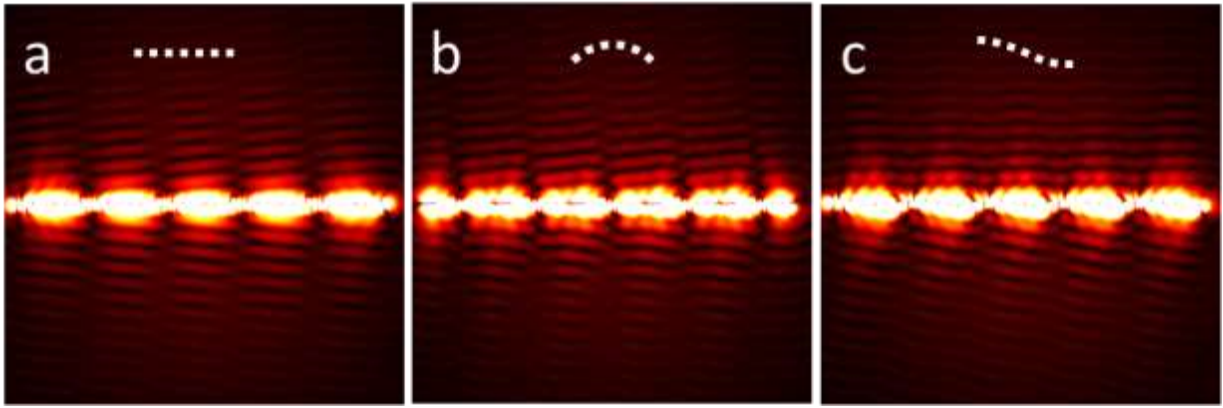

**Figure 3.** Analytical simulations. Results highlighting the different phase gradients that can be created with a phased line of V-antennas. (a) 1D phase gradient, denoted by the white dashed line, resulting in symmetric surface plasmon wavefronts, for a period of $\Gamma = 3.0 \ \mu m$ and left circularly polarized light, where $\gamma_1 = \gamma_2 = 5.8°$. (b) 2D phase gradient, as shown in Fig. 2c, resulting in nearly symmetric wavefronts (phase gradient in $y$ cancels out along the period) but with a slight focusing effect on the bottom side of the metastructure due to the curvature of the gradient. (c) 2D phase gradient, as shown in Fig. 2d, which is monotonic decreasing and well-approximated by a line, giving rise to asymmetric wavefronts on either side of the array of V-nanoslits, where the wavefront angles are given by Eq. 1 and 2 as $\gamma_{top} = 2.4°$ and $\gamma_{bottom} = 9.2°$.

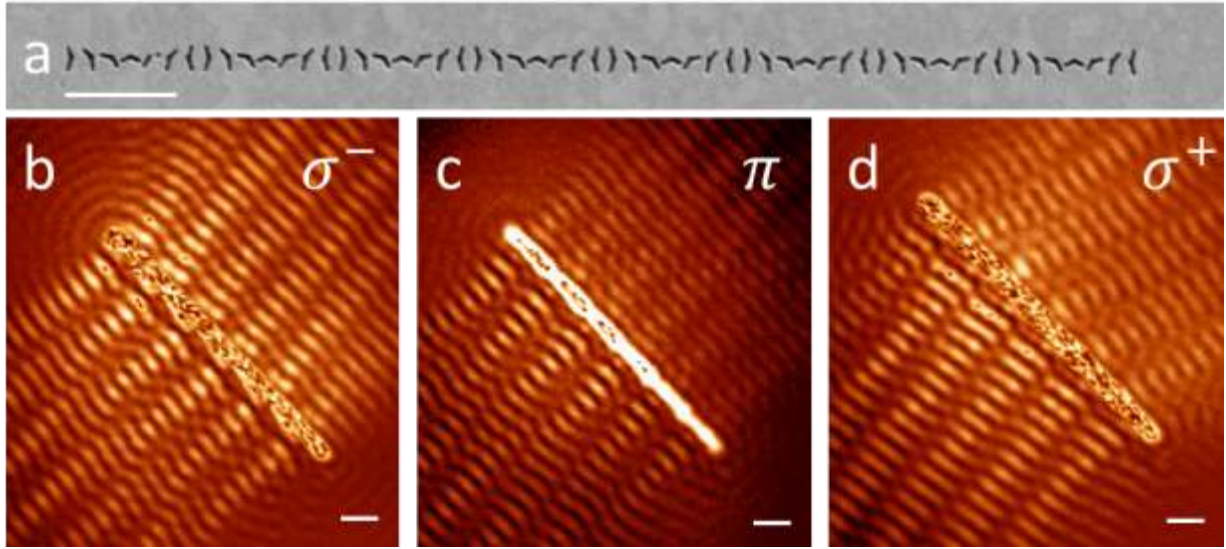

**Figure 4.** NSOM data for the focusing arrangement. a) Scanning electron micrograph highlighting the focusing 2D gradient. b) NSOM data for left circularly polarized light with a focusing effect visible on the left/underside of the structure and wavefront angles dictated by the Equations 1 and 2 with $\Delta y = 0$. c) NSOM data for linearly polarized light, which serves to further highlight the curved wavefronts due to focusing—it is more apparent for linearly polarized light because there is no phase gradient, so the geometrical effect is enhanced. d) NSOM data for right circularly polarized light. Note that the wavefront angles become equal and opposite to those in (b) because the handedness of the light has been reversed.

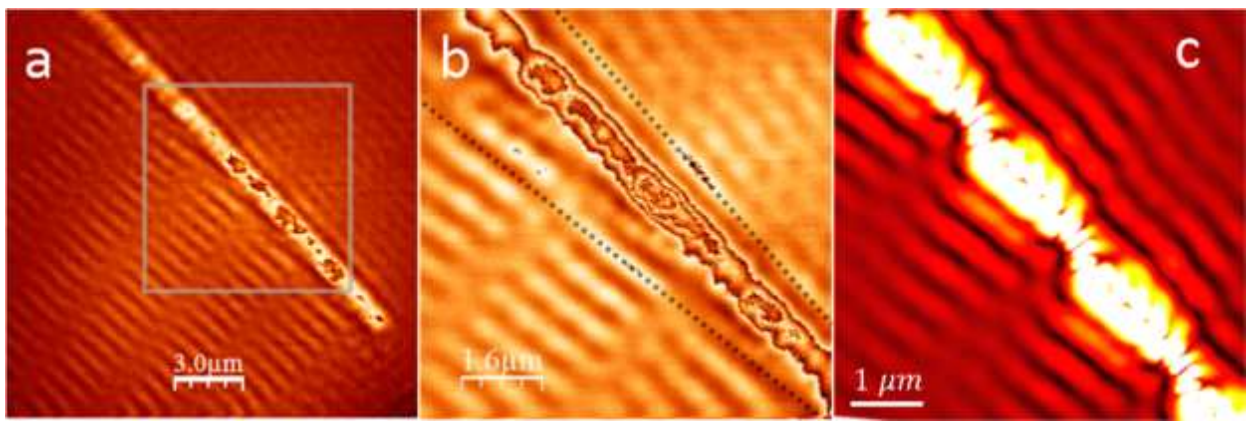

**Figure 5.** NSOM data for V-antennas that produce asymmetric wavefront angles. a) Full-view NSOM data for period $\Gamma = 3.0\ \mu m$. b) Zoom in for the area denoted by the grey box in (a). The wavefront angle asymmetry is clearly visible on either side of the structure, matching well with theory for an antisymmetric mode and symmetric mode separation of $\approx \lambda_{SPP}/6.8$, deduced from the wavefront angles that are measured $\gamma_1 = 9.2°$ and $\gamma_2 = 2.4°$. c) Rotated, 3D rendering of (b), used to highlight the asymmetry.

**Table of Contents Graphic and Synopsis**

We utilize bimodal antennas to impose a 2D phase gradient and observe asymmetric surface plasmon polariton wavefronts.

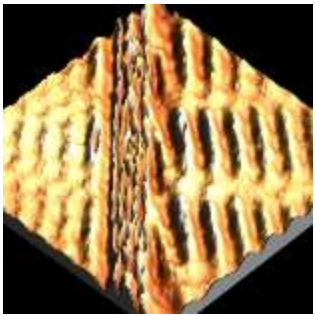